\documentclass[preprint,aps]{revtex4}
\usepackage{graphicx}
\begin{document}

\title{Stabilization of microtubules due to microtubule-associated 
proteins: A simple model}

\author{Bindu S. Govindan}
\affiliation{Applied Biosciences Center,
Virginia Polytechnic Institute  and State University,
Blacksburg, VA 24061-0356, U. S. A.}

\author{William B. Spillman, Jr.}

\affiliation{Applied Biosciences Center and Department of Physics,
Virginia Polytechnic Institute  and State University,
Blacksburg, VA 24061, U. S. A.
}

\date{\today}

\begin{abstract}

A theoretical model of stabilization of a microtubule assembly due to microtubule-associated-proteins(MAP) 
is presented. MAPs are assumed to bind to the microtubule filaments, thus preventing their disintegration 
following hydrolysis and 
enhancing further polymerization. Using  mean-field rate equations and explicit numerical simulations, we 
show that the density of MAP (number of MAP per tubulin in the microtubule) has to exceed a critical value
$\rho_{c}$ to stabilize the structure against de-polymerization.  At lower densities $\rho < \rho_{c}$, 
the microtubule population consists mostly of short polymers with exponentially decaying length distribution, 
whereas at $\rho > \rho_{c}$ the average length increases linearly with time and the microtubules 
ultimately extend to the cell boundary. Using experimentally measured values of various parameters, 
the critical ratio of MAP to tubulin required for unlimited growth is seen to be of the order of 1:100 or even smaller.
\end{abstract}

\maketitle

\section{Introduction}
Microtubules are hollow, cylindrical polymers which perform several
important functions inside the cell. The persistence length of a typical
microtubule is of the order of mm \cite{ref.1},  so they behave as rigid rods inside cells (linear dimension $\sim1-10\mu m$). 
The basic monomer unit of a microtubule is a heterodimer of $\alpha$ and $\beta$-tubulin. Within a microtubule, tubulin heterodimers 
are arranged head-to- tail to 
form linear protofilaments, usually 13 per tube. The length of a tubulin dimer is nearly 8 nm, and the total diameter 
of a microtubule 
is about 25 nm. Heterodimers are oriented with their $\beta$-tubulin monomer pointing toward the plus end, or fast 
growing end, of the microtubule. The minus ends are at the microtubule-organizing centers such as centrosomes in animal
cells.

Both $\alpha$ and $\beta$-tubulin binds GTP. Following polymerization, the $\beta$-tubulin in the dimer hydrolyses 
its bound GTP to GDP \cite {ref.2}. The hydrolyzed GDP-tubulin (which we shall refer to as D-tubulin, while
the un-hydrolyzed version shall be called T-tubulin) does not polymerize, and 
the process is irreversible inside the microtubule.
When the advancing hydrolysis front reaches the growing end of a microtubule, it starts de-polymerizing from the end and the 
protofilaments start peeling off, releasing
the GDP-tubulin into solution. Outside a microtubule, reverse hydrolysis takes place
and polymerization events start all over again.
Thus the microtubule constantly switches between phases of 
growth and shrinkage, and this behavior (first observed in {\it in vitro} experiments) is called {\it dynamical instability}\cite{ref.3,ref.4}. When a microtubule makes a transition from growing to shrinking state, it is said to undergo {\it catastrophe} and the reverse transition 
is called {\it rescue}. The relative rates of catastrophe and rescue, combined with
the velocities of growth and shrinkage determine the character of a given
population of microtubules \cite{ref.5,ref.6,ref.7}.

The intrinsic microtubule dynamics observed {\it in vitro} is modified inside the cell by interaction with 
cellular factors that stabilize or destabilize microtubules, and which operate in both spatially and temporally 
specific ways to generate different types of microtubule assemblies during the cell cycle. Proteins that modulate 
microtubule dynamics are called microtubule-associated proteins (MAP). Several MAPs have been 
identified, including MAP1, MAP2 and tau in neurons and MAP4 in non-neuronal cells \cite {ref.8,ref.9,ref.10,ref.11}. 
MAPs regulate the microtubule dynamics through one or more of the following mechanisms:
suppressing catastrophe rate, increasing rescue rate, promoting polymerization and preventing de-polymerization \cite{ref.12}.

Recent experimental studies show that, inside cells, 
dynamic instability of microtubules is observed predominantly near the cell margin. Deep inside the 
cytoplasm, microtubules were observed to be in a state of almost persistent growth with very few catastrophe events. 
Direct experimental 
measurements showed a marked difference between the rates of catastrophe in the cell interior 
($\nu_c \approx 0.005 s^{-1}$) and periphery ($\nu_c \approx 0.08s^{-1}$)[13].
These observations suggest that in the cell interior, microtubule growth is sustained by cellular  factors which are 
presumably absent or rare near  the cell boundary. Our principal aim in this paper is to develop a simple
mathematical model to study how the microtubule-associated proteins affect the dynamics of microtubule assembly. 

It is convenient to outline the main results of this paper at this point. By analyzing the rate equations associated 
with microtubule dynamics under a mean-field treatment of MAP density, we show that the microtubule assembly can be in two 
different phases depending on the proportion of 
MAP to tubulin (henceforth referred to as the MAP density) inside the microtubule. Above a critical density $\rho_{c}$, 
individual microtubules exist in a phase of unlimited growth, while below this density, growth is limited and the 
average length is finite. We show that $\rho_{c}=p_H/(p_g+p_H)$, where $p_H$ and $p_g$ are rates of hydrolysis and 
growth respectively. The rates of catastrophe and rescue are related to $\rho$ through the relations $\nu_{c}=p_{H}(1-\rho)$ 
and $\nu_{r}=p_{s}\rho$ where $p_{s}$ is the rate of shrinkage. Numerical simulations show that the fluctuations in MAP
distribution do not signficantly affect the predictions based on mean-field analysis.
We compare our results to experimental observations on the
effects of MAP on microtubules, and find that there is qualitative agreement.

This paper is arranged as follows. In the next section, we briefly discuss the salient features of our model and its
limitations.In Sec. 3, we write down explicitly the rate equations that govern the dynamics of the microtubule
assembly. We derive expressions for the critical MAP density, the time evolution of the average length of microtubules 
in the unlimited growth regime as well as the expressions for the rates of catastrophe and rescue in our model. 
In Sec.4, we present results of numerical simulations of our model which are in excellent agreement with the 
analytical predictions. In Sec.5, we summarise our results and outline the main conclusions that emerges out of this study.

\section{The Model}
In this model, we assume that microtubules are thin rigid rods nucleating from the microtubule-organizing center, which is very often the centrosome(Fig.1). The three-dimensional 
structure of microtubules is ignored in our model. The centrosome contains nucleation sites for a microtubule, and nucleation takes place 
at vacant sites at a rate $\nu$. The microtubule grows by adding tubulin subunits at a rate $p_{g}$, and tubulin units inside 
the microtubule are stochastically hydrolyzed at a rate $p_{H}$. A microtubule will start de-polymerizing when the T-tubulin at its 
growing end is hydrolyzed to D-tubulin
\cite {ref.3,ref.14,ref.15}. The shrinkage takes place at a rate $p_{s}$, which is the probability per unit time to lose a 
subunit. It is generally assumed that the de-polymerization process will continue along the length
of the microtubule, until the shrinking end encounters a patch of T-tubulin in the
interior. In our model, however, we assume that a microtubule in shortening state will continue to shrink (irrespective of 
the state of tubulin in the interior) until the de-polymerizing end encounters a MAP, which are present along the 
microtubule. This simplification helps focus on the role of MAPs in the microtubule dynamics. We further 
assume that MAPs also assist in the polymerization process by acting as polymerization sites along a microtubule.
For further simplification, we assume the presence of an excess of free tubulin in solution, so that the dynamics of the 
tubulin concentration can be largely ignored. We also assume that the nucleation sites in the 
centrosome are spaced sufficiently far apart so that the effective interaction between microtubules through local depletion 
and  enhancement of concentration of free tubulin is insignificant. A schematic illustration of the dynamics of our model
is given in Fig.2, and a schematic illustration of the various rates of the model is shown in Fig.3.

It is helpful to note the typical experimental values of the various parameters in our model. We refer to recent
experimental work \cite{ref.13} for most of the
parameter values. Nucleation of microtubules in the centrosome was seen to occur 
at a rate of $\nu\simeq$ 5-6/min {\it in vivo} \cite {ref.13}. 
The velocity of growth of growing microtubules observed is $v_g\approx 20 \mu m/min$. The rate of addition 
of tubulin dimers ($p_g$) is determined from $v_{g}$ as follows. A  tubulin dimer has length $\approx 8 nm$. Since there are 
13 protofilaments of T-dimers in a single microtubule, each tubulin contributes a length of $\delta \approx 0.6 nm$ to the microtubule. 
Thus, the rate of addition of tubulin dimers would be $p_g= v_{g}/\delta\simeq 0.55\times 10^{3}s^{-1}$. Using similar 
reasoning, the rate of loss of tubulin can be calculated from the measured velocity of shrinkage in shortening 
microtubules. This turns out be $p_s=v_s/\delta\approx 0.82\times 10^{3}s^{-1}$. 

The rate of hydrolysis is a difficult parameter to measure, and is not as well known as the other rates.
However, a wide range of values, 0.25-25/min has been reported
\cite {ref.2,ref.14,ref.16,ref.17}. 
The typical values of the density $\rho$ of MAP  used in injection experiments are estimated to be of the order 
$10^{-2}$ \cite{ref.18,ref.19}.

\section{Mean-Field Rate Equations}

In this section, we analyze the mean-field rate equations for the time evolution of the probability 
distribution of the lengths of microtubules. Since our basic aim is to study the character of an assembly of microtubules, 
we resort to a statistical description. The most appropriate quantity to study in this perspective is the length distribution 
$p(l,t)$, which is the fraction of nucleation sites which has a microtubule of  
length $l$ at time $t$. The presence of MAP is taken into account through a mean-field density variable $\rho$, whose local
fluctuations in space is ignored (The effect of these fluctuations are explicitly studied in numerical simulations presented
in Sec.4). The MAPs change the microtubule dynamics by modifying the rates of shrinkage and hydrolysis, as seen below.
It is also convenient to classify the entire set of microtubules into two: growing and 
shrinking microtubules (as was done in \cite{ref.15}). Consequently, the complete length distribution $p(l,t)$ can also be split as follows.

\begin{equation}
p(l,t)=p_{+}(l,t)+p_{-}(l,t)
\label{eq:SPLIT}
\end{equation}

\noindent
where $p_{+}(l,t)$ is the fraction of microtubule in growing state at any time $t$, and
$p_{-}(l,t)$ is the fraction in shrinking state at time $t$.
The rate of change of $p_{+}(l,t)$ is described by the 
following equation (for $l\geq 2$):

\begin{equation}
\frac{\partial p_{+}(l,t)}{\partial t}=p_{+}(l-1,t)p_g +p_{-}(l+1,t)\rho p_{s}
-p_{+}(l,t)[p_{g}+p_{H}(1-\rho)]~~~~; ~~~l\geq 2
\label{eq:RATE1}
\end{equation}

The significance of each term easily follows from the model description in Sec.2. The first two terms represent 
{\it gain} events, where a microtubule of length $l$ in growing state is created at time $t$. In particular, 
the second term
describes the transition between a microtubule in a shrinking state to a growing state, and so 

\begin{equation}
\nu_{r}=\rho p_{s}
\label{eq:RES-RATE}
\end{equation}

\noindent
is the rate of {\it rescue} in our model. We note that in our model $\nu_{r}\to 0$ as $\rho\to 0$, which is in
agreement with the assumptions of our model. (However, in experiments, microtubules have been 
observed to undergo rescue under {\it in vitro} conditions in the absence of MAP. Our assumption 
here is that the MAP significantly increases the rescue frequency, which would otherwise be negligible). The last two terms represent the {\it loss} events 
where a microtubule of length $l$ either adds a tubulin to increase its  length, or undergoes hydrolysis to transform into 
shrinking state. The latter is an event of {\it catastrophe}, and so the corresponding rate in our model is

\begin{equation}
\nu_{c}=p_{H}(1-\rho). 
\label{eq:CAT-RATE}
\end{equation}

\noindent
Both the rates of catastrophe and rescue are parameters which are measured in experiments.
We also note that for microtubules of unit length, the gain event in $p_{+}(1,t)$ is nucleation, rather than 
polymerization, so that its equation is

\begin{equation}
\frac{\partial p_{+}(1,t)}{\partial t}=\nu R(t)+p_{-}(2,t)\rho p_{s}-p_{+}(1,t)[p_{g}+p_{H}(1-\rho)]
\label{eq:RATE1+}
\end{equation}

\noindent
where $R(t)=1-\sum_{l=1}^{\infty}[p_{+}(l,t)+p_{-}(l,t)]$ is the fraction of vacant sites in the lattice 
at time $t$. The equation for $p_{-}(l,t)$ is,

\begin{equation}
\frac{\partial p_{-}(l,t)}{\partial t}=p_{+}(l,t)p_{H}(1-\rho)+p_{-}(l+1,t)p_{s}(1-\rho)-p_{-}(l,t)p_{s}~~~;~~l\geq 1
\label{eq:RATE2}
\end{equation}

\noindent
The equations Eq.(\ref{eq:RATE1}) and Eq.(\ref{eq:RATE2}) are supplemented by the 
boundary condition for $p(l,t)$, which is,

\begin{equation}
lim_{l\to\infty}p(l,t)=0
\label{eq:BOUNDARY}
\end{equation}

\noindent
We now proceed to look for the solution for these coupled set of rate equations. As a first step, we look for 
the steady state solution by putting the time derivatives to zero in Eq.(\ref{eq:RATE1}) and Eq.(\ref{eq:RATE2}).

\subsection{The steady state:}

\vspace{1cm}
In the steady state, it is reasonable to assume that both the distributions $p_{+}(l)$ and $p_{-}(l)$ (the time 
dependence has been dropped since we are looking at steady state solutions now) have the same dependence on $l$. 
Further, it is easily seen from the equations that the steady state solution has an exponential form, which is 
easily determined by substituting the trial functions

\begin{equation}
p_{+}(l)=Ae^{-\lambda l}~~~;~~~p_{-}(l)=\gamma p_{+}(l)
\label{eq:TRIAL}
\end{equation}

\noindent
in the steady state equations. In the trial solution, $\lambda$ and $\gamma$ are two unknown constants which need to 
be determined, and $A$ is the normalization factor. Upon substitution 
in Eq.(\ref{eq:TRIAL}) we find the following equations for the unknown constants $\lambda$ and $\gamma$.

\[
p_g[e^{\lambda}-1]+\gamma p_{s}\rho e^{-\lambda}=p_{H}(1-\rho)=p_{s}
\gamma[1-(1-\rho)e^{-\lambda})]
\]

\noindent
from which we determine the unknown constants $\lambda$ and $\gamma$ as follows.

\begin{equation}
\lambda=log[(1+\frac{p_H}{p_g})(1-\rho)]~~~~~;~~\gamma=\frac{p_{g}+p_{H}}{p_{s}}(1-\rho)
\label{eq:LAMBDA}
\end{equation}

\noindent
Clearly, from the boundary condition in Eq(\ref{eq:BOUNDARY}), the solution is meaningful 
only if $\lambda >0$, and 
this implies that $\rho <\rho_{c}$, where

\begin{equation}
\rho_{c}=\frac{p_H}{p_g+p_H} 
\label{eq:RHOC}
\end{equation}

\noindent
It turns out that $\rho_{c}$ is a critical value of MAP density, below which growth is bounded and the length 
distribution of microtubule is stationary. The typical values of $\rho_{c}$ can be easily estimated using
the known values of $p_{H}$ and $p_{g}$. For example, keeping $p_{g}\simeq 0.55\times 10^{3}s^{-1}$ and changing
$p_{H}$ from $\simeq$ 3/min to 300/min (which are close to {\it in vivo} estimates) gives 
$\rho_{c}$ in the interval $\sim 10^{-4}-10^{-2}$. This suggests that a very small MAP to tubulin ratio might
be enough to stabilize microtubules against depolymerization.

The steady state solution discussed above is not valid when $\rho\geq \rho_c$, and so the solution
for the rate equations has to be found explicitly for this case. 

\subsection{The unbounded growth regime}

When $\rho > \rho_{c}$, growth becomes unbounded. Although the exact solution of the evolution equations in this
regime is hard to find, it is helpful to note that under usual experimental conditions the growth and shrinkage
rates far exceed the hydrolysis rate, i.e.,$p_{H}$ is small in comparison with $p_{g}$ and $p_{s}$. If we further
assume that we work in the regime where the MAP density is high, i.e., $\rho\simeq 1$, then all the terms in the rate
equations which contain $p_{H}$ can be neglected in comparison with the other terms.
Further analytic progress is now possible by converting the finite difference equations into differential equations 
(in the limit $l\gg 1$). This is done by 
formally expanding $p_{\pm}(l,t)$ in its derivatives (see the appendix for a discussion on this point) as the series,
$p_{\pm}(l\pm 1,t)=p_{\pm}(l,t)\pm \frac{\partial p_{\pm}}{\partial l}+\frac{1}{2}\frac{\partial^2p_{\pm}}{\partial l^2}$.
After substitution, we arrive at the following differential equation for $p_{-}(l,t)$.

\begin{equation}
\frac{\partial p_{-}(l,t)}{\partial t}=\frac {p_{s}}{2}(1-\rho)\frac{\partial^2 p_{-}(l,t)}{\partial l^2}+p_s(1-\rho)\frac{\partial p_{-}(l,t)}{\partial l}- p_{-}(l,t)p_{s}\rho
\label{eq:CONT-MINUS}
\end{equation}

Let us now define the re-scaled time $\tau=\rho p_{s} t$ and the new variables $D=\frac{1-\rho}{2\rho}$ 
and $v=\frac{1-\rho}{\rho}$, in terms of which Eq.(\ref{eq:CONT-MINUS}) is re-cast as 

\begin{equation}
\frac{\partial p_{-}(l,\tau)}{\partial t}=D\frac{\partial^2 p_{-}(l,\tau)}{\partial l^2}+v \frac{\partial p_{-}(l,\tau)}{\partial l}-p_{-}(l,\tau) 
\end{equation}

\noindent
This equation is easily solved by  substituting $p_{-}(l,\tau)=e^{-\tau}g(l,\tau)$, upon which we find that $g(l,\tau)$ satisfies the 
equation $\frac{\partial g}{\partial \tau}=D\frac{\partial^2 g}{\partial l^2}+v\frac{\partial g}{\partial l}$,
which is the well-known equation for diffusion of a particle in one dimension, with a drift term. The solution 
of this equation (apart from the trivial solution $g(l,\tau)=0$) has the form(up to a proportionality factor),

\begin{equation}
 g(l,\tau)=\frac{1}{\sqrt{D\tau}}e^{-\frac{(l+v\tau)^2}{4D\tau}} 
\end{equation}

\noindent
However, this solution is not physically reasonable since the mean length decreases with time as $\langle l\rangle=-v\tau$. 
It follows that in this limit, the solution for Eq.(\ref{eq:CONT-MINUS}) is the trivial solution i.e, $p_{-}(l,t)\simeq 0$. 
The equation for $p_{+}(l,t)$ in the same limit is,

\begin{equation}
\frac{\partial p_{+}(l,\tau)}{\partial t}=\frac{p_g}{2}\frac{\partial^2 p_{+}(l,\tau)}{\partial l^2}-p_g \frac{\partial p_{+}(l,\tau)}{\partial l}
\label{eq:CONT-PLUS} 
\end{equation}

\noindent
which, again, is the diffusion equation with a drift term, whose solution is (up to a constant)

\begin{equation}
p_{+}(l,t)=\frac{1}{\sqrt{t}}e^{-\frac{(l-p_{g} t)^2}{2p_{g} t}}
\label{eq:CONT-PLUS-SOLUTION}
\end{equation}

\noindent

We see that in the limits $p_{H}\ll p_{s}$ and $\rho\simeq 1$, the length distribution of growing microtubules 
is a Gaussian at all times, and the mean value increases linearly with time as $\langle l \rangle\approx p_{g}t$. Furthermore, in this limit, 
almost all the microtubules are in the growing state. 

Although it is possible at this stage to look for a perturbative solution of the evolution 
equations Eq.(\ref{eq:CONT-MINUS}) and Eq.(\ref{eq:CONT-PLUS}) with $\frac{p_H}{p_s}$ as the pertrubation 
parameter, we have not attempted this here. Rather, in the next subsection, we determine the mean length
as a function of time exactly, and shows that it agrees with the prediction of Eq.(\ref{eq:CONT-PLUS-SOLUTION})
in the limit $p_{H}\ll p_{s}$.

\vspace{1cm}

\subsection{The average length of microtubules}

In order to study the time evolution of the average length in the unbounded growth regime, it is convenient to define the following quantities:

\begin{equation}
Q_{+}(t)=\sum_{l=1}^{\infty}p_{+}(l,t)
\end{equation}

\noindent
which is the fraction of microtubule in growing state, irrespective  of length, and 

\begin{equation}
Q_{-}(t)=\sum_{l=1}^{\infty}p_{-}(l,t)
\end{equation}

\noindent
is the fraction of microtubule in shrinking state, irrespective of length. The time evolution of $Q_{+}(t)$  can be found by performing a sum of Eq. (\ref{eq:RATE1}) over all $l\geq 2$ and adding Eq.(\ref{eq:RATE1+}).The result is 

\begin{equation}
\frac{dQ_{+}(t)}{dt}=\nu R(t)-p_{H}(1-\rho)Q_{+}(t)+\rho p_{s}[Q_{-}(t)-p_{-}(1,t)]
\label{eq:Q+}
\end{equation}

\noindent
For $Q_{-}(t)$, we perform a sum over Eq.(\ref{eq:RATE2}) over all $l\geq 1$, and find that 

\begin{equation}
\frac{dQ_{-}(t)}{dt}=p_{H}(1-\rho)Q_{+}(t)+
p_{s}(1-\rho)[Q_{-}(t)-p_{-}(1,t)]-p_{s}Q_{-}(t)
\label{eq:Q-}
\end{equation}

\noindent
where $R(t)=1-Q_{+}(t)-Q_{-}(t)$. The average length of microtubule at time $t$ is defined as 

\begin{equation}
\langle l\rangle(t)=\sum_{l=1}^{\infty}l[p_{+}(l,t)+p_{-}(l,t)]
\end{equation}

\noindent
The rate of change of $\langle l\rangle$ with time may be expressed in terms of $Q_{+}(t)$ and $Q_{-}(t)$, 
since each growing microtubule grows by  an amount $p_{g}$ and each shrinking microtubule shrinks by an amount $p_{s}$ per unit time.
It follows that

\begin{equation}
\frac{d\langle l\rangle }{dt}=Q_{+}(t)p_{g}-Q_{-}(t)p_{s}
\label{eq:DLDT}
\end{equation}

To calculate $Q_{+}(t)$ and $Q_{-}(t)$ from Eq.(\ref{eq:Q+}) and Eq.(\ref{eq:Q-}), let us first assume that 
$p_{-}(1,t)\ll Q_{-}(t)$. This assumption is valid in the unbounded growth regime, where the mean length 
increases with time, and 
hence the peak of the distribution is expected to shift toward larger and larger times (note the shifting
Gaussian we found in the last section). The result is a closed set of equations for $Q_{+}(t)$ and $Q_{-}(t)$, 
which has the form

\begin{equation}
\frac{dQ_{+}(t)}{dt}=\nu (1-Q_{+}-Q_{-})-\alpha Q_{+}+\beta Q_{-}
\end{equation}

and

\begin{equation}
\frac{dQ_{-}(t)}{dt}=\alpha Q_{-}-\beta Q_{+}
\end{equation}

\noindent
where $\alpha=p_{H}(1-\rho)$ and $\beta=\rho p_{s}$. We take the initial condition to be 
$Q_{+}(0)=\nu$ and $Q_{-}(0)=0$. The general solution for the above set of equations is

\begin{equation}
Q_{-}(t)=\frac{\alpha}{(\alpha+\beta)}+\frac{\alpha\nu(1-(\alpha+\beta))}{(\alpha+\beta)(\alpha+\beta-\nu)}
e^{-(\alpha+\beta)t}-\frac{\alpha(1-\nu)}{(\alpha+\beta-\nu)}e^{-\nu t}
\label{eq:EARLYQ-}
\end{equation}

\noindent
and
\begin{equation}
Q_{+}(t)=\frac{\beta}{\alpha+\beta}+\frac{\alpha\nu(1-(\alpha+\beta))}{(\alpha+\beta)(\alpha+\beta-\nu)}
e^{-(\alpha+\beta)t}+\frac{\alpha(1-\nu)(\nu-\beta)}{(\alpha+\beta-\nu)}e^{-\nu t}
\label{earlyQ+}
\end{equation}

\noindent
We note that the exponential terms vanish at sufficiently late times, and so the asymptotic forms of $Q_{+}(t)$ and
$Q_{-}(t)$ are,

\begin{eqnarray}
Q_{+}(t)\to \frac{p_{s}\rho}{p_{s}\rho+p_{H}(1-\rho)}~~~~~;~~~t\to\infty\nonumber\\
Q_{-}(t)\to \frac{p_{H}(1-\rho)}{p_{s}\rho+p_{H}(1-\rho)}~~~~;~~~t\to\infty
\label{eq:Q+-}
\end{eqnarray}

\noindent
Thus, even in the unbounded growth phase, a finite fraction of
microtubule will be in the shrinking state. After substituting  Eq.(\ref{eq:Q+-}) in Eq.(\ref{eq:DLDT}), we 
find that 

\begin{equation}
\frac{d\langle l\rangle }{dt}\approx \frac{p_{s}[p_{g}\rho-p_{H}(1-\rho)]}{\rho p_{s}+p_{H}(1-\rho)}~~~~;~~~t\to\infty
\end{equation}

For unbounded growth, we need $\frac{d\langle l\rangle }{dt} >0$, 
for which the condition $\rho > \rho_{c}$, where $\rho_{c}$ is defined in Eq.(\ref{eq:RHOC}). This calculation confirms that $\rho_{c}$ is the critical MAP density above which microtubules exist in a phase
of sustained growth. In the limit $p_{H}\ll p_{s}$, we see that $\langle l\rangle\simeq p_{g}t$, which is one of the
results we had derived in the last subsection.

\vspace{1cm}

\subsection{ Comparison with experiments}

\vspace{1cm}

At this stage, it is worthwhile to compare our predictions (although based on a very simple model) 
with experimental observations regarding the effect of MAPs on microtubules\cite{ref.18,ref.19}. We 
specifically refer to 
experiments by Dhamodharan and Wadsworth \cite{ref.19} where MAPs were
microinjected into BSC-1 cells and their effects on growth of microtubules, in particular, the changes
in catstrophe and rescue frequencies, were studied. When 0.1 mg/ml of MAP-2 was added to the cells, the proportion
of MAP to tubulin in the polymerized state was estimated to be nearly 1:25, or $\rho\simeq 0.04$. Increasing
the MAP concentration 10-fold was observed to decrease the catastrophe frequency from 0.05 ($\pm 0.02$)$s^{-1}$
to 0.02 ($\pm 0.003$)$s^{-1}$, i.e., by a factor of nearly 0.4 after neglecting the experimental error.
According to Eq. (\ref{eq:CAT-RATE}), increasing $\rho$ by a factor of 10, from 0.04 to 0.4, would decrease the
catastrophe frequency by a factor of 0.625, which is in resonably good agreement with the experimental
value. Furthermore, according to Eq. (\ref{eq:CAT-RATE}), the catastrophe rate in the absence of
MAP is simply the hydrolysis rate $p_{H}$, which may be calculated from the same equation using 
experimental data for $\nu_{c}$ and $\rho$. The data for $\rho=0.4$ (which has greater accuracy) gives
$p_{H}\approx 0.033 s^{-1}$, which is, again, in reasonably good agreement with the experimental
value of 0.043 ($\pm 0.01$) $s^{-1}$. The effects of MAP on the rescue frequency is more subtle, since
rescue would occur in cells even in the absence of MAP, which is not taken into account in our model. 

\section{NUMERICAL SIMULATIONS}

In this section, we discuss the results of our numerical simulations of the microtubule
assembly. Since our model neglects any effective interaction between growth of neighboring microtubules due
to local depletion of ligand concentration, the geometry of the system, and the spacing between individual microtubules 
is unimportant in the simulation. For simplicity, the cell interior is imagined as a rectangular box, where the 
base forms the substrate for nucleation of microtubules \cite {ref.15}. 
The microtubules grow in the direction perpendicular to the base until they hit the boundary. 

The next step in the simulation procedure is to divide the box into a grid array of dimensions $L\times L\times H$, where
$L$ is the length of the side of the substrate and $H$ is the height. As we had mentioned before, the lattice spacing
in the substrate is unimportant as long as it is sufficiently large that direct interaction between neighboring microtubules
could be ignored (henceforth, we shall always assume that this condition is satisfied). 
The lattice spacing in the direction of growth ($z$-axis) is the length of one tubulin
dimer, which, after taking into account the number of protofilaments, is equivalent to $\delta z\simeq 0.6 nm$. 
For the time evolution, we choose Monte Carlo step as equivalent to $\delta t\simeq 10^{-3}s$. With this timescale,
the probability of a microtubule growing by one unit per MC step is nearly 0.55, and the probability that a tubulin unit
is lost in one MC step is nearly 0.82, from the rates discussed in Sec. 2. In other words, the time scale has been 
chosen such that growth and shrinkage of microtubules are stochastic events, in accordance with the rate equations.
For the lattice size, we chose $L=100$ and $H=500$.

The tubulin density {\it in polymerized state} at each point in space is represented
using a density index $\phi({\bf r},z,t)$, where ${\bf r}$ is the substrate co-ordinate, $z$ is the
position in the direction of growth and $t$ is time in units of MC steps. 
At $t=0$, we start with a substrate free of microtubule, i.e., $\phi({\bf r},z,0)=0$. We do not explicitly
keep track of the density of free tubulin in the solution. The density
index $\phi({\bf r},z,t)$ takes three values: $\phi({\bf r},z,t)=+1$ represents T-tubulin,
$\phi({\bf r},z,t)=-1$ represents D-tubulin and $\phi({\bf r},z,t)=0$ implies that the site is
empty. A string of $\phi({\bf r},z,t)\neq 0$ at any ${\bf r}$, for $1\leq z\leq l$ is 
a microtubule of length $l$. For $z=1$, $\phi({\bf r},z,t)$ changes from 0 to $+1$ with probability $\nu$ per
MC step at all ${\bf r}$ and $t$. This is the nucleation event. For any $z >1$, this transition takes
place with probability $p_g$, which is a polymerization event where a new T-tubulin dimer 
is added to the microtubule. The location of MAP in the lattice is specified through another density index 
$\chi({\bf r},t)$, which takes the value 1 or 0 depending on whether there is a MAP molecule at the lattice site
({\bf r},z) at time t. At all times, we distribute the MAPs at random throughout the lattice with mean density $\rho$.
We do not take into account the diffusion of MAPs in solution explicitly. Rather, when microtubules grow 
from the substrate, they assimilate the MAPs along their length and release them randomly such that the mean density 
remains $\rho$ at all times. We assume here that the binding and release of MAP take place over time scales much 
less than one 1MC step in  the simulation (which is equivalent to $10^{-3}$ seconds), so that, at each time step, 
the microtubule assembly  has a different MAP configuration.
 
In all our simulations, the rates of growth and shrinkage  were fixed as $p_{g}=0.55$ and  $p_{s}=0.82$, which are
the experimental values quoted in Sec.2.  For reasons we shall shortly discuss, the hydrolysis rate was fixed at 
a value higher than the typical estimates, i.e, we chose $p_{H}=0.5$ (which would translate
into a hydrolysis rate of $\approx 3\times 10^{4}$/minute). We also chose a high nucleation rate ($\nu=1$), 
for two reasons: to increase the occupancy of the lattice so as to improve the statistics, and to reduce the 
duration of the early-time regime of growth (where the exponential factors in Eq..(\ref{eq:EARLYQ-}) and 
Eq..(\ref{earlyQ+}) would ne non-negligible). The chosen values of $p_{H}$ and $p_{g}$ gives a convenient value
of the critical MAP density, $\rho_{c}\simeq 0.47$ (which was the primary reason for choosing a high $p_{H}$).

We measured the length distribution and average length of microtubule assembly for three values of 
$\rho$, 0.1, 0.3 and 0.6. We chose the first two values to be below $\rho_{c}$, where we expect bounded growth 
(steady state distribution) and  the last value is above $\rho_{c}$, where
growth is expected to be unbounded. The results for probability distribution of lengths are given in Figs. 4-6, and 
the result for time evolution of the average length in the unbounded regime is displayed in Fig.7.
In Fig. 4 and 5, we observe that, as predicted by the rate equation analysis, the solution is exponential, but
the measured value of $\lambda$ differs from the mean-field prediction. This is presumably due to the spatial fluctuations
in the MAP density, which is ignored in the mean-field analysis. In Fig.6, we find that
the length distribution of microtubules in the unbounded regime closely resembles a Gaussian function with a mean value
shifting with time. This is in accordance with the results of rate equation analysis in Sec. 3 (although in the 
simulations $p_{H}$ and $p_{s}$ are of the same order of magnitude). In Fig.7, we confirm that the mean length in the
unbounded regime follows a linear dependence on time, again in agreement with the analysis in Sec.3.

\section{SUMMARY AND CONCLUSIONS}

In this work, we have presented a simple theoretical model of stabilization of
microtubules due to microtubule-associated proteins. We have shown that when the density of 
MAP is reduced below a critical value, the length distribution of the microtubule assembly undergoes a 
significant change. Whereas in one phase (at high MAP density) 
the microtubules extend throughout the cell, in the other phase they have an exponentially
decaying length distribution with a finite average length. It is interesting to note that this
transition is similar to the transition at mitosis in real cells, where phosphorylation of MAPs by
agents such as the metaphase promoting factor (MPF)\cite {ref.10} leads to a weakening of their binding to the 
microtubules, and a consequent change in length distribution. We have also shown that the critical ratio
of MAP to polymerized tubulin required for a transition to the 'cell-spanning' phase can be as low as 1:1000 under
{\it in vivo} conditions. Thus, our conclusion is that even a very small concentration of MAP can sustain unlimited growth 
of microtubules, which is in qualitative agreement with experiments \cite{ref.18}.

There are still many open questions regarding the effect of MAPs on microtubule dynamics in cells. In fact, MAPs
which inhibit polymerization have also been identified recently \cite {ref.11}. It would be interesting to include such MAPs
also in our model and look for the consequences of competition between the effects of two kinds of MAP. 
Moreover, our prediction of the existence of a critical MAP density 
which differentiates between two distinct growth regimes for microtubules could be verified in future
experiments. Lastly, our model could be still improved by including effects of local fluctuations in the concentrations
of MAP and tubulin. In particular, it would be interesting to ask in the context of such an improved model whether
there exists subpopulations of microtubules in one growth regime, while the over-all growth behavior of the whole
population is in the other regime. These are some of the questions we would like to address in the future.

\section{ACKNOWLEDGMENTS}

The authors would like to gratefully acknowledge the Carilion Biomedical Institute for 
providing the funding which allowed this work to be carried out. One of us (B.G) would like 
to thank Prof. R. A. Walker for helpful discussions.

\begin{center}
{\bf APPENDIX}
\end{center}

In this appendix, we outline the scheme by which the difference equations Eq..(\ref{eq:RATE1}) and Eq..(\ref{eq:RATE2}) 
are converted to partial differential equations Eq..(\ref{eq:CONT-MINUS}) and Eq..(\ref{eq:CONT-PLUS}). 
We first define the first derivative in a symmetric way by taking the average over right and left derivatives.

\[
\frac{\partial p_{\pm}}{\partial l}\equiv \frac{[p_{\pm}(l+1)-p_{\pm}(l)]+[p_{\pm}(l)-p_{\pm}(l-1)]}{2}
\]

We note that the second and third terms cancel each other, so we arrive at

\[
\frac{\partial p_{\pm}}{\partial l}\equiv \frac{p_{\pm}(l+1)-p_{\pm}(l-1)}{2}
\]

For the second derivative, we use the familiar Euler discretization scheme, i.e,

\[
\frac{\partial^2 p_{\pm}}{\partial l^2}\equiv p_{\pm}(l+1)+p_{\pm}(l-1)-2p_{\pm}(l)
\]

After solving these two equations together, we arrive at the expression

\[
p_{\pm}(l\pm 1,t)=p_{\pm}(l,t)\pm \frac{\partial p_{\pm}}{\partial l}+\frac{1}{2}\frac{\partial^2p_{\pm}}{\partial l^2}
\]

\newpage

\begin{figure}
\includegraphics{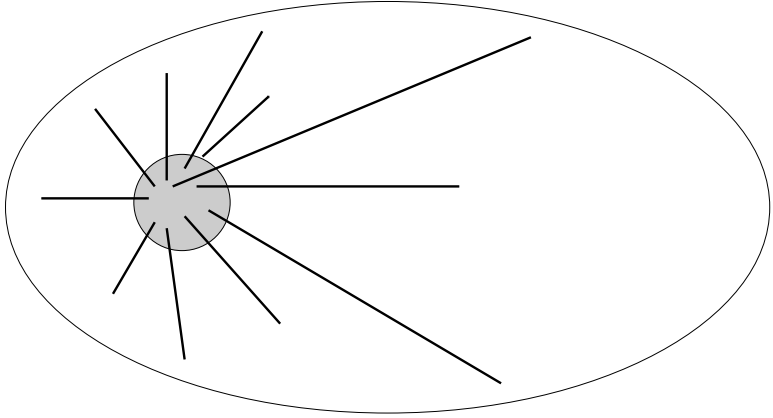}
\caption{ \small A schematic illustration of microtubules inside the cell. The filled circle is the microtubule-organizing
center, which is very often the centrosome.}
\end{figure}

\begin{figure}
\includegraphics{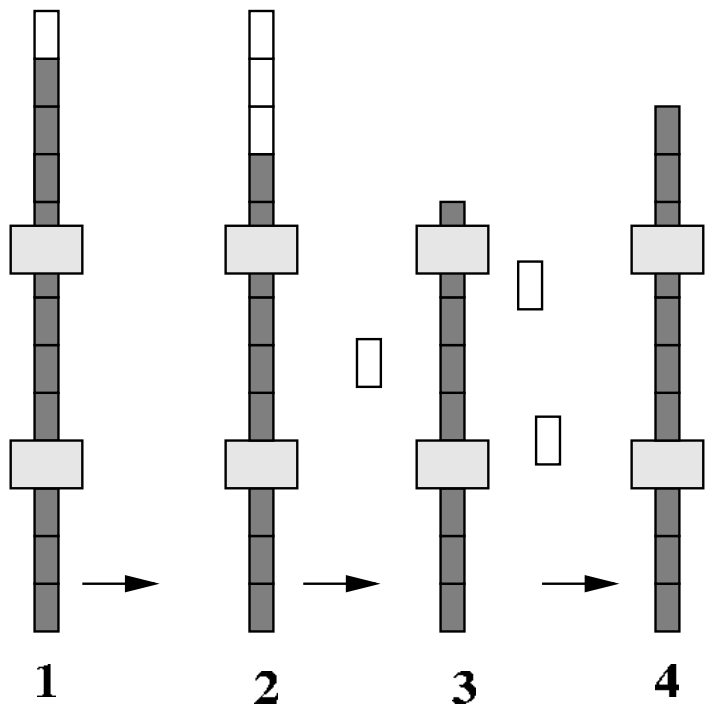}
\caption{\small A schematic illustration of the dynamics of our model. The dark squares are the T-tubulin units and
the white squares are the hydrolysed D-tubulin units. The large filled squares represent the microtubule-associated
proteins. In (1), hydrolysis of the microtubule has started from the top end, which spreads to the bottom (2), and
untimately leads to the D-tubulin units falling off the polymer. But the de-polymerization is arrested in
(4) by the MAP, which promotes further polymerization.}
\end{figure}

\begin{figure}
\includegraphics{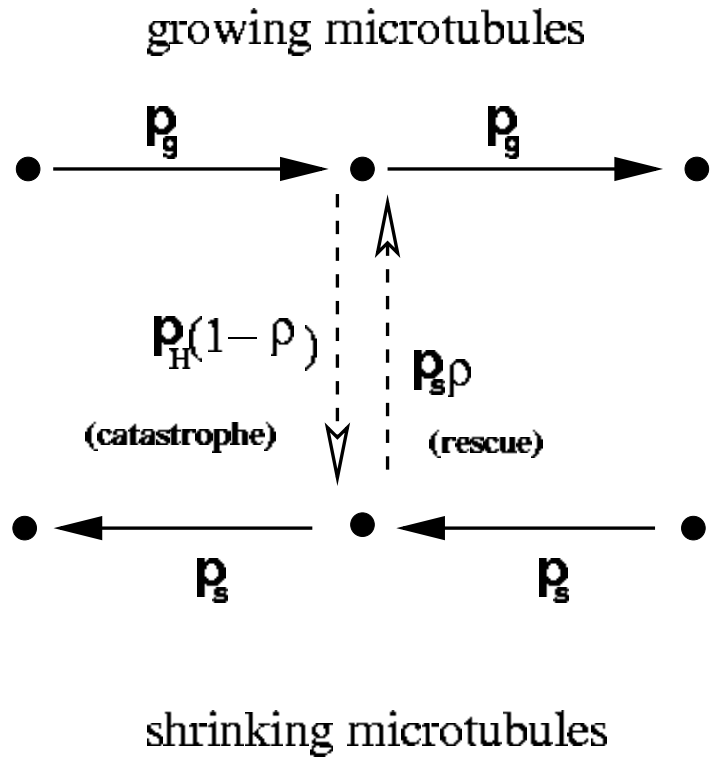}
\caption{ \small A schematic illustration of the various rates in the microtubule dynamics. Mirotubules make the transition
between growing and shrinking states with rates which depend on the MAP density.}
\end{figure}

\begin{figure}
\includegraphics{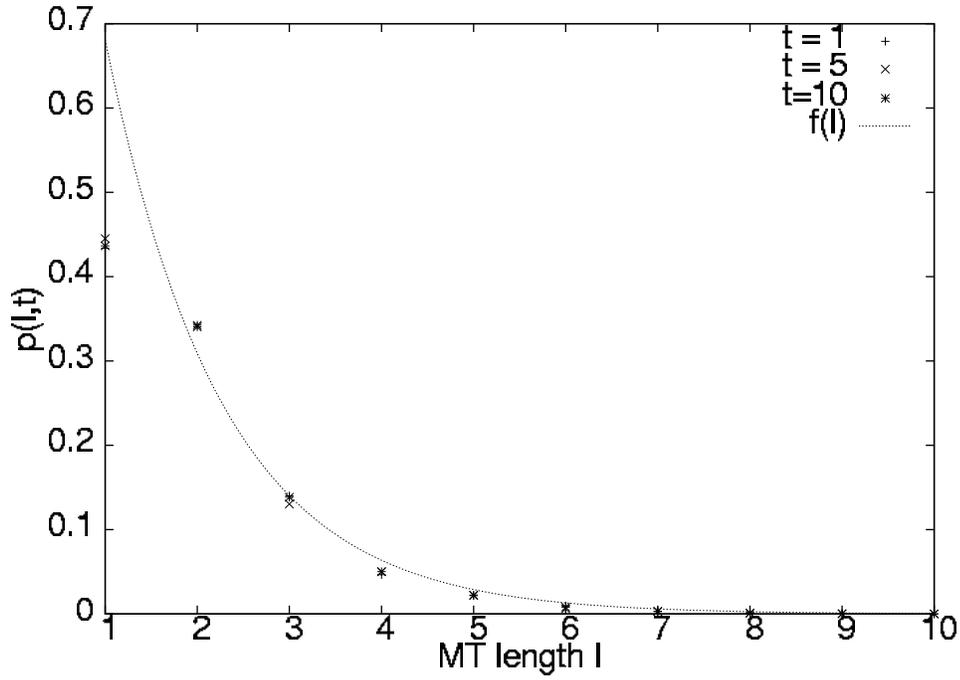}
\vspace{-12.0cm}
\caption
{ Probability distribution of microtubule length when $\rho=0.1$. 
We observe bounded growth with
exponential distribution of lengths as predicted by the rate equations. The unit 
of time in this plot
is one second. $f(l)\propto e^{-\lambda l}$ is a fit function where $\lambda\approx 0.79$, 
which
differs somewhat from the mean-field prediction of Eq..(\ref{eq:LAMBDA}), which 
gives $\lambda\simeq 0.54$.}
\end{figure}

\begin{figure}
\includegraphics{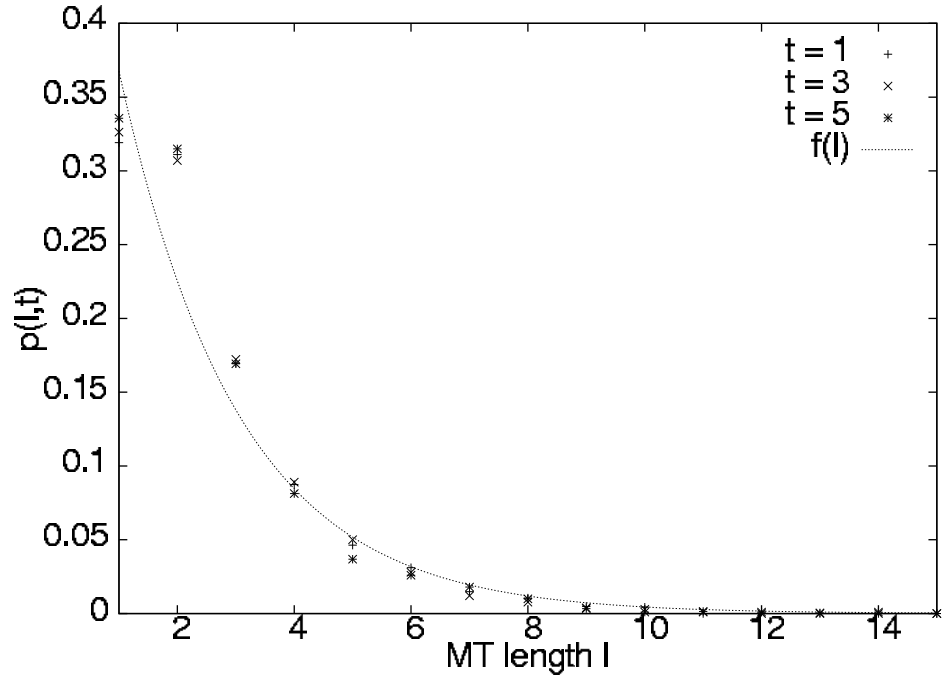}
\vspace{-10.cm}
\caption{\small The figure shows the probability distribution of microtubule lengths when $\rho=0.3$. 
In this case, growth is again bounded,
but the length scales reached by the microtubule is longer than before. The unit of time in this plot
is one second. $f(l)\propto e^{-\lambda l}$ is a fit function, with 
$\lambda\approx 0.49$. The mean-field prediction for $\lambda$ given by Eq
.(\ref{eq:LAMBDA}) is nearly 0.28.}
\end{figure}

\begin{figure}
\includegraphics{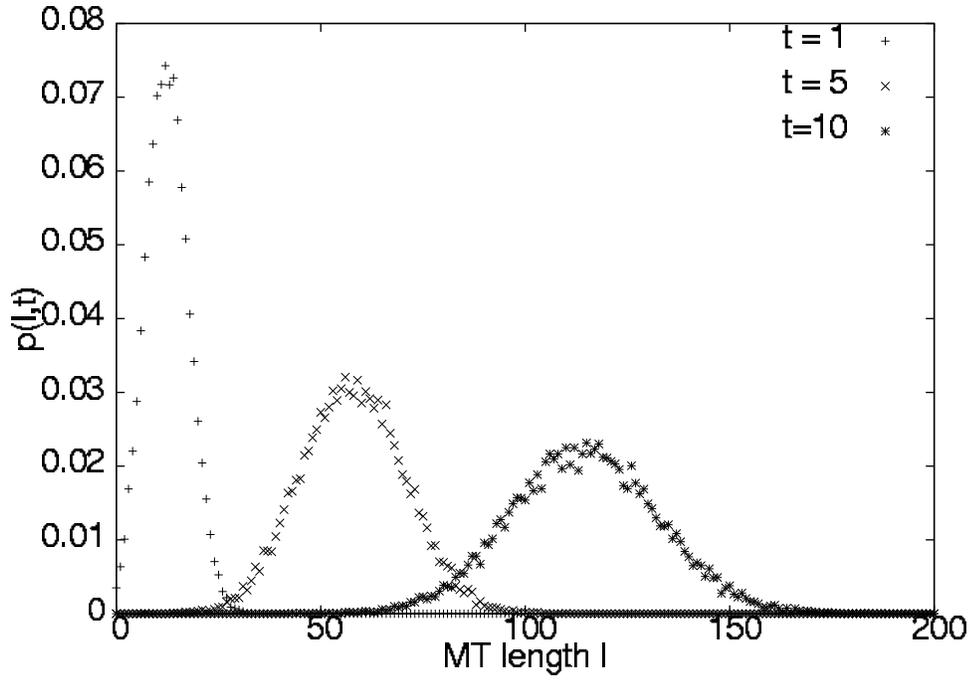}
\vspace{-10.0cm}
\caption{\small The figure shows the probability distribution of microtubule lengths when $\rho=0.6$ at three different
times. The growth is unbounded since $\rho > \rho_{c}$. 
The peak of the distribution shifts to the right with increasing time, which 
shows that the average length increases with time. The unit of time in this figure
is one second and length is measured in units of $\delta$ = 0.6nm.}
\end{figure}

\newpage
\begin{figure}
\includegraphics{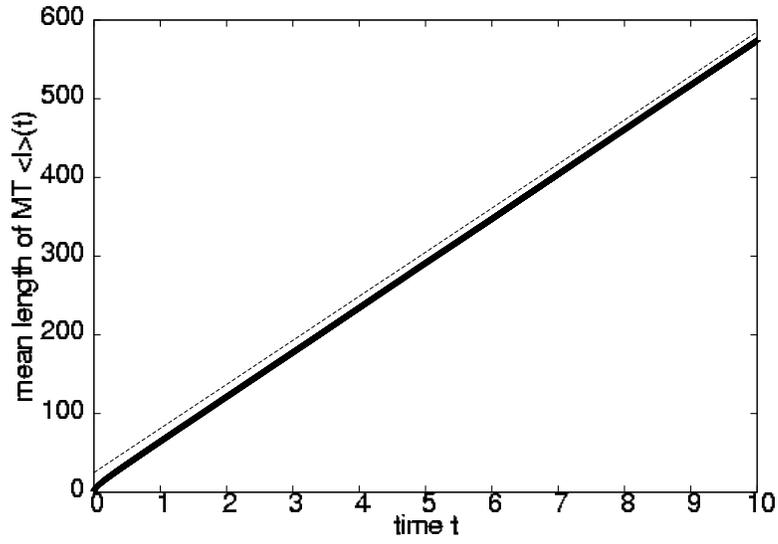}
\vspace{-5.0cm}
\caption{\small The plot shows the average length of the microtubule assembly increasing linearly with time for 
$\rho=0.6$, which is above $\rho_{c}$. The straight line is a linear fit. Time is measured in
seconds and length is measured in units of $\delta$ = 0.6nm.}
\end{figure}

\end{document}